\documentclass{rspublic}

\usepackage{graphicx}

\usepackage{moreverb,ulem,color}

\begin{document}


\title[Robustness of Estimators of Long-Range Dependence and
  Self-Similarity]{Robustness of Estimators of Long-Range Dependence and
  Self-Similarity under non-Gaussianity}

\author[C. L. E. Franzke, T. Graves, N. W. Watkins, R. B. Gramacy and
  C. Hughes]{Christian L. E. Franzke$^{A1}$,Timothy Graves$^{A2}$, Nicholas
  W. Watkins$^{A1}$, Robert B. Gramacy$^{A2, A3}$ and Cecilia Hughes$^{A1}$}

\affiliation{$^{A1}$ British Antarctic Survey, Cambridge, UK\\[1.0ex] $^{A2}$
  Statistical Laboratory, University of Cambridge, Cambridge, UK\\[1.0ex]
  $^{A3}$ Booth School of Business, The University of Chicago, Chicago, USA}

\label{firstpage}

\maketitle

\begin{abstract}{Memory, Paradigmatic Models, Multiplicative Noise}
Long-range dependence and non-Gaussianity are ubiquitous in many natural
systems like ecosystems, biological systems and climate. However, it is not
always appreciated that both phenomena may occur together in natural systems
and that self-similarity in a system can be a superposition of both
phenomena. These features, which are common in complex systems, impact the
attribution of trends and the occurrence and clustering of extremes. The risk
assessment of systems with these properties will lead to different outcomes
(e.g. return periods) than the more common assumption of independence of
extremes.

Two paradigmatic models are discussed which can simultaneously account for
long-range dependence and non-Gaussianity: Autoregressive Fractional
Integrated Moving Average (ARFIMA) and Linear Fractional Stable Motion
(LFSM). Statistical properties of estimators for long-range dependence and
self-similarity are critically assessed. It is found that the most popular
estimators can be biased in the presence of important features of many natural
systems like trends and multiplicative noise. Also the long-range dependence
and non-Gaussianity of two typical natural time series are discussed.
\end{abstract}

\section{Introduction} 
Meteorology has been both a seedbed and a testbed for many advances in the
mathematics  of complex systems; the former being seen in its contribution of
the Lorenz model to nonlinear science (Lorenz 1963), while the latter is
exemplified by ensemble forecasting. This parallel evolution of theory and
practise has identified important problems, e.g., the need to model the effect
of extra, {\it noise-like}, degrees of freedom on deterministic low
dimensional dynamics. Other well established paradigms such as Hasselmann's
stochastic model (Hasselmann 1976), have highlighted the importance of {\it
red noise} to mathematical climatology.

However, since the pioneering work of the late Benoit Mandelbrot, increasing
attention is paid to two self-similar (or ``fractal") aspects of time series:
long-range dependence (LRD) in time, and the spatial counterpart of LRD, heavy
tailed probability  distributions in amplitude. First identified in hydrology,
and since studied in research areas as diverse as biology, telecommunications,
social networks, econometrics and the climate system, LRD is characterised by
its low frequency singular behaviour, the so-called $1/f$ power spectrum. When
present in a signal the correlations captured by LRD will both hamper the
identification  and the attribution of deterministic trends (e.g. Franzke
2010), and impede the quantification of their significance.  As was the case
in the original context of hydrology, the presence of LRD in climate time
series has been intensely debated, and it is still sometimes ascribed to
transient effects, calibration issues or other forms of nonstationarity
(Maraun et al. 2004, Rust et al. 2008).

LRD is also of practical significance. By making systems subject to
longer-lived fluctuations (Beran 1994) it changes the information available to
make predictions about the state of a complex system. It impacts the
occurrence and clustering of extremes (Bunde et al. 2005, Bogachev et
al. 2008, Kropp and Schellnhuber 2010) which are important for risk
assessments and mitigation strategies, e.g. insurance pricing and flood
defence. The traditional assumption is that extremes are independent
events. But there is growing evidence for clustering of extreme extra-tropical
storms (e.g. Mailier et al. 2006) and precipitation events (e.g. 2007 United
Kingdom floods). This clustering of extremes will lead to higher return
periods of extreme events than the more common assumption of
independence. Recently, detrended fluctuation analysis (DFA), a tool
originally developed to detect LRD, has found application in the prediction of
dangerous bifurcations in dynamical systems such as climate ``tipping points"
(Livina and Lenton 2007, Lenton et al. 2011, Sieber and Thompson 2011). Our
contribution to this volume is focused both on i) the problem of accurately
estimating LRD in the presence of other signal elements, and ii) the
complicating effects of heavy-tailed amplitude distributions, when they are
present.

A process is long range dependent when the prediction of its next state
depends on the whole of its past. An imprint of this dependence structure is
that the covariance $\ r(k)=Cov(X(k),X(0))$ decays slowly, as$\ k \rightarrow
\infty$, so that
\begin{equation} 
\sum_{k=0}^{\infty} | r(k) | = \infty. 
\label{eq:longrange} 
\end{equation} 
This slow decay of the covariances means that the values of the process
\textit{X} are strongly dependent over long periods of time. This contrasts
with the more familiar short-range dependent system where $
\sum_{k=0}^{\infty} | r(k) | = C < \infty $. In a short-range dependent
process the next state only depends on the current state and the recent
past. The archetype of a short-range dependent process is a first order Markov
process where the next state depends only on the present state and is
conditionally independent of past states. See Beran (1994) for more details.

Long-range dependence of a system is characterised by the parameter $d$ in the
statistics literature and sometimes by Mandelbrot's Joseph parameter $J$, for
example in the physics literature. Temporal long-range dependence has been
detected in the water level of the river Nile (Hurst 1951, Hurst et
al. 1965). It was observed that the range of values grows as $\tau^J$, where
$J$ refers to the Joseph exponent (e.g. Mandelbrot 2001, p. 157) and $\tau$ to
the time period under consideration. The growth of range was anomalously large
compared to that in the familiar paradigms of random walks and Brownian
motion, which have both played a central role in understanding diffusion. In
the random walk, the variance grows as the square root of time $ \tau^{1/2} $
(Einstein 1905). The subsequent theoretical explanation of the difference
between the observed anomalous diffusion and the Brownian motion paradigm came
first through the use of a self-similar process, in particular via the study
of fractional Brownian motion (fBm; Mandelbrot \& Wallis 1968). Here,
self-similarity means that the statistical properties on all scales are
similar. This property is controlled by the self-similarity parameter $H$. A
stochastic process $S$ is self-similar when a rescaling of time by a factor $
\lambda $ leads to a rescaling of the amplitude of the process $S$ by $
\lambda^H $. Thus, a process is said to be $H$-self-similar when the following
relation holds (Embrechts and Maejima 2002):
\begin{equation}
 S(\lambda t) \stackrel{d}{\sim} \lambda^H S(t). 
\label{eq:selfsim}
\end{equation}
where $ \stackrel{d}{\sim} $ denotes equality in distribution. It is not
always appreciated, and is sometimes confused in the literature, that two
different aspects of a time series can contribute to its self-similarity
$H$. The first is LRD, which is synonymous with persistence, referred to by
Mandelbrot and Wallis (1968) as the {\it Joseph} effect. Persistent systems
exhibit longevity by having a tendency to maintain the way they have been
recently. Examples are heat waves and drought conditions. The second source of
self-similarity identified by Mandelbrot, the so-called {\it Noah} effect is
the property that change in a system can be rather large and can occur very
abruptly i.e. time series drawn from systems can exhibit sharp
discontinuities, e.g., Earth quakes.

It may at first seem odd that both phenomena occur simultaneously in natural
systems because they pull in opposite directions. On reflection, taken
together, however, we see that  the two effects capture the facts that
coherent structures in nature are real and that they can emerge, change or
even vanish very quickly. The archetypal model of the Noah effect is
non-Gaussian jumps whose complementary cumulative distribution function decays
with a power law in size $s$, $p(s) \sim s^{- \alpha}$. Here, $\alpha$ denotes
the stability exponent (referred to as tail exponent in the statistics
literature). Thus, the self-similarity parameter is
\begin{equation}
H = J -\frac{1}{2} + \frac {1} {\alpha} = d + \frac {1} {\alpha}, \quad H \in
(0,1). 
\label{EQ:1}
\end{equation}
The distinction between $H$ and $J = d + \frac {1} {2} $ is important because
not all observed time series have Gaussian fluctuations. As such one may find
that popular diagnostics, as shown below, may be insensitive to heavy tails,
measuring J and not H. However, in many situations both heavy tails and the
clustering of extremes caused by LRD are very important and both contribute to
the value of H. This makes it necessary to be able to measure both H and J
independently.

Many estimators for the LRD exponent $d$ have thus been developed, and are
widely used in the respective literatures. Much of what has been rigorously
established about the  estimators, however, is for a particular LRD Gaussian
model: fBm, which is also $H$ self-similar and has a particularly simple
relation between $H$ and $J$, i.e. $ H = J + \frac {1} {2}-\frac{1}{2}=J $ ($
\alpha =2 $ for Gaussian increments). Most observed time series depart from
the assumptions of fBm in some way or another, so it is important to
critically evaluate how sensitive the estimators are to deviations from
fractional Gaussian noise (fGn). fBm is a random walk with long-range
correlated increments and the value of $J$ in such $H$-self-similar walks is
directly connected to the presence of LRD in their increments.

In this study we critically re-evaluate 4 popular methods for measuring $d$
and two for measuring $H$: variable bandwidth estimator (Schmittbuhl et
al. 1995) and wavelets (WL; Stoev and Taqqu 1995), Detrended Fluctuation
Analysis (DFA; Peng et al. 1994), the Re-Scaled Range analysis ($R/S$; Hurst
1951, Mandelbrot 2001), exact Whittle estimator (Shimotsu and Phillips 2005,
2006) and semi-parametric estimators (in this study we will focus on the
Geweke-Porter-Hudak (GPH) estimator; Geweke \& Porter-Hudak 1983;
semi-parametric estimators are also described in Bardet et al. 1998, Robinson
1995a, 1995b, Hurvich et al. 2005). This is a not a complete set of estimators
for the self-similarity or long-range dependence parameter (other estimators,
which are well established in the statistics community include FEXP (Hurvich
et al. 2002)). Here we want to focus on some of the most popular estimators in
the physics and geosciences communities ($R/S$: Tuck and Hovde 1999, Price and
Newman 2001, Ogurtsov, 2004, Scipioni et al, 2008, Ghil et al. 2011, DFA: Chen
et al. 2002, Bunde et al. 2005, Bashan et al. 2008, Ghil et al (2011), GPH:
Vyushin and Kushner 2009, Huybers and Curry 2006, Franzke 2010, $\mathrm{VB}$:
Escorcia-Garcia et al. 2009, WL: Stoev and Taqqu 2005, Stoev et al. 2005).

We show scatter plots which compare the imposed parameters for a number of
trials with the values detected by the above methods. These have the advantage
of visually indicating which methods measure the complete self-similarity
exponent $H$, and which, instead, measure long range dependence via $d$. For
evaluating the performance of these estimators we employ 2 well-established
paradigmatic time series models: A self-similar process (Linear Fractional
Stable Motion (LFSM); Samorodnitsky \& Taqqu 1994) and an asymptotically
self-similar process (Autoregressive Fractional Integrated Moving Average
(ARFIMA); e.g. Beran 1994).

In most natural systems trends are common properties, so we systematically
examine the performance of the various estimators when a linear trend is
superposed on  each time series. This is an important issue because in
practice it is not always easy to remove trends. That they typically manifest
themselves in higher moments presents further challenges still. Trends are the
hardest to deal with because long-range dependent processes can produce
apparent trends over rather long periods of time (e.g. Franzke
2010). Therefore, it can be difficult to decide if a trend is due to external
forcing or due to finite time series length.

In section 2 we present the two paradigmatic time series models in more detail
and discuss various special cases. In section 3 we introduce the estimators of
long-range dependence and self-similarity. We test their utility in section
4. In section 5 we discuss the LRD and self-similarity properties of two
exemplary time series from nature and in section 6 we provide conclusions.
\begin{figure} 
\begin{center}
\includegraphics[width=30pc]{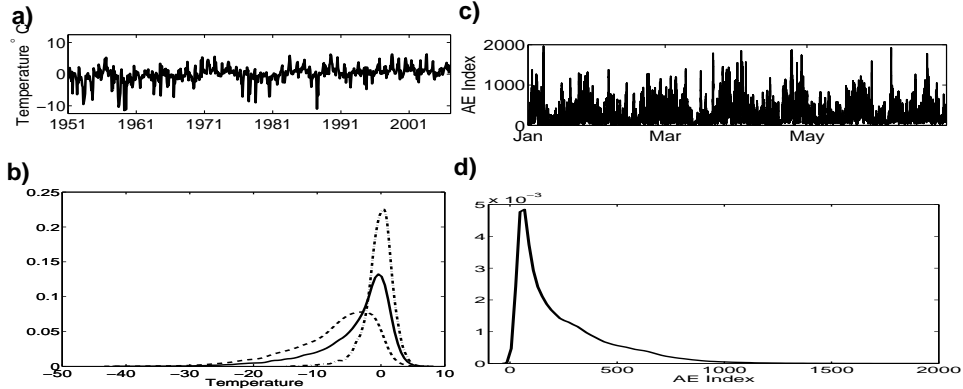}
\caption{\label{FARADAYAE}a) Faraday-Vernadsky station temperature time series
  in degree Celsius (annual cycle is subtracted) (01.01.1951 - 28.02.2007), b)
  Probability density function of the Faraday-Vernadsky station temperature
  time series estimated by a Kernel density estimator: Solid line: whole
  year; Dashed line: Cold season data May through September; Dashed-Dotted
  line: Warm season data November through March. c) Auroral electrojet index
  from January through June 2006 in nanotesla and d) Probability density
  function of the Auroral electrojet index.}
\end{center}
\end{figure}
\section{Paradigmatic Models of Natural Time Series}
\subsection{Natural Time Series Examples}
The most prominent and widely used paradigmatic LRD time series model,
especially in the physics literature, is fGn. However, most natural time
series  are not strictly Gaussian and many are actually highly
non-Gaussian. To illustrate this point we discuss two time series from
nature. In Fig.~\ref{FARADAYAE}a we display monthly mean temperature from the
Antarctic station Faraday-Vernadsky (Turner et al. 2004, Franzke 2010). It is
easy to see that the raw time series is highly non-Gaussian. This is confirmed
by the probability distribution functions (PDFs) plotted in
Fig. \ref{FARADAYAE}b. There is also a striking seasonal signal visible in the
skewness of the PDFs. However, it must to be noted that polar temperature time
series are much more non-Gaussian than mid-latitude ones which are typically
nearly Gaussian. Our second exemplary time series is the auroral electrojet
(AE) index (Fig. \ref{FARADAYAE}, e.g., Davis \& Sugiura 1966, Watkins
2002). The AE index is derived from 1 minute resolution time series from 12
high latitude magnetometers. Reflecting the intermittent nature of the
ionospheric and solar wind processes which influence it, the AE index is seen
to be spiky and strongly non-Gaussian (Fig. \ref{FARADAYAE}d). Clearly, many
natural time series are non-Gaussian.

\subsection{Paradigmatic Models}
While fGn and fBm are Gaussian, they are still useful idealised paradigmatic
models for understanding many observed phenomena. A paradigmatic model is an
idealised framework which captures some properties of observed time series,
though not all. Most paradigmatic models allow analytical work, although at
the expense of an over-idealisation of the physical or biological
phenomena. There is a need for better paradigmatic models of observed
phenomena which allow simultaneously non-Gaussian statistics and LRD. Such
models are needed e.g., for hypothesis testing and time series
simulation. Thus, we briefly discuss two paradigmatic time series models which
exhibit LRD and non-Gaussianity. The widely used fGn and fBm are special cases
in the Gaussian limit of these more general models, thereby retaining some of
the analytical tractability of these models.

In this study we use two classes of processes with symmetric$\ \alpha $-stable
(S$ \alpha$S) distributions. They are the LFSM and the ARFIMA model with S$
\alpha$S innovations. Stoev and Taqqu (2004) present efficient methods for the
simulation of these processes using the Fast Fourier Transform.  
\subsubsection{Linear Fractional Stable Motion}
LFSM is a model which exhibits LRD and heavy tails at the same time. It is an
extension to the simpler Brownian random walk model. It links individual heavy
tailed jumps by means of a retarded memory kernel. It can be represented by a
stochastic process$\ X_{H,\alpha}= \left\{ X_{H,\alpha}(t), t \in \mathbb{R}
\right\}$,\footnote{This is the traditional way of defining LFSM, though $H$
has been defined in (\ref{EQ:1}) as the dependent and $d$   and $\alpha$ as
the independent variables.} which is defined by the following stochastic
integral
\begin{equation}
 X_{H,\alpha}(t) = C_{H,\alpha}^{-1} \int_{\mathbb{R}} \left(
(t-s)^{H-\frac{1}{\alpha}}_{+} - (-s)^{H-\frac{1}{\alpha}}_{+} \right)
dL_{\alpha}(s),
\label{eq:stoch}
\end{equation}
where $ 0<H<1$, $ \alpha \in (0,2)$, and where $ L_{\alpha} =
\left\{L_{\alpha}(s),s \in \mathbb{R} \right\}$ is a standard symmetric $
\alpha$-stable (S$ \alpha$S) Levy process. The process $\ X_{H,\alpha}$ is
called LFSM and is self-similar with self-similarity parameter H. Thus, it
satisfies relation (\ref{eq:selfsim}) and has stationary increments. The
parameter $ \alpha $ controls the tails of the distribution of an S$ \alpha$S
random variable $ \xi$, that is,
\begin{equation}
\mathbb{P} \left(|\xi|\geq x \right) \sim x^{-\alpha}, \mbox{as } x
\rightarrow  \infty.
\label{eq:tails}
\end{equation}
The greater the value of$\ \alpha$, the lower the probability of extreme
fluctuations of the S$ \alpha$S process$\  X_{H,\alpha}$. We recommend the
introductions by Taqqu (2003) and Mercik et al. (2003) for detailed
expositions.

\subsubsection{Fractional Brownian Motion (fBm)}
In the Gaussian case we have that $\alpha=2$ and the LFSM process
(\ref{eq:stoch}) reduces to fBm. In this case the self-similarity parameter
$H$ will always equal the Joseph exponent $J$, thus, $ H = J = d + \frac {1}
{2} $.
\subsubsection{Ordinary L\'{e}vy Motion (oLm)}
In the memory-less case we have $d=0$ and $ 0< \alpha < 2$. In this case
equation (\ref{eq:tails}) holds and the tails of the distribution decay
according to a power-law and consequently $ L_{\alpha}$ has infinite variance
and is called ordinary L\'{e}vy motion. This encapsulates the Noah effect. In
this case the self-similarity exponent is $ H=\frac{1}{\alpha} $. 
\subsubsection{Autoregressive Fractional Integrated Moving Average}
Another paradigmatic model exhibiting LRD with heavy-tailed fluctuations is
ARFIMA, which has the added ability of exhibiting short-range dependent
behaviour. This model, denoted ARFIMA(\textit{p,d,q}), $ p,q \in \mathbb N$,
extends the usual Autoregressive Integrated Moving Average
ARIMA(\textit{p,d,q}) models, in which $d$ takes integer values (Box \&
Jenkins 1970). An ARIMA model is written as:
\begin{equation}
\Phi(B)(1-B)^dX_t = \Psi(B) \epsilon_t,
\end{equation}
where \textit{B} denotes the back shift operator defined by $ BX_t=X_{t-1}$,$
B^2X_t=X_{t-2}$... The polynomials $ \Phi$ and $ \Psi $ are defined as $
\Phi(x)=1-\sum_{j=1}^{p} a_j x^j $ and $ \Psi(x)=1+\sum_{j=1}^{q} b_j
x^j $, where \textit{p} and \textit{q} are integers. The innovations
$\epsilon_t~(t=1,2...)$ are usually independent and identically distributed
(iid) normal variables with zero expectation and variance $
\sigma_{\epsilon}^2 $, but can also be $\alpha$-stable distributed. The widely
used autoregressive process of first order (AR(1)) is a special case with
$p=1$, $d=0$ and $q=0$. Typically $d$ is an integer. In the case that $d$ is a
fractional real number, \textit{X(t)} is an ARFIMA(\textit{p,d,q}) process
with $ -\frac{1}{2} < d < \frac{1}{2} $ and exhibits LRD. In the Gaussian case
ARFIMA is stationary. An ARFIMA $(0,d,0)$ is asymptotically equivalent to fBm
(Taqqu 2003). An ARFIMA$(p,d,q)$ with $p>0$ and $q>0$ is long-range dependent
but not self-similar. However, it is asymptotically self-similar for long time
scales.

\section{Estimators of Long-Range Dependence and Self-Similarity}
Here we briefly describe the four estimators of the self-similarity and LRD
parameters used in this study.
\subsection{Variable Bandwidth}
The variable bandwidth ($\mathrm{VB}$) method (Schmittbuhl et al. 1995) is a
technique for estimating the self-similarity exponent, $H$, from a time series
$x(t)$. The time series of length $T$ is divided into windows or `bands' of
width $r$.  The $\mathrm{VB}$ method can deploy two different algorithms to
estimate $H$. (i) The standard deviation of the time series, $\sigma(r)$, is
computed in each band; or (ii) the difference between the maximum and minimum
values in each band, $\epsilon(r)$, is computed. Then $\sigma(r)$ and
$\epsilon(r)$ are averaged over all the possible bands by varying the origin
at fixed $r$ 
\begin{equation}
\mathrm{VB}_w(r)=\frac{1}{L_r}\sum_{i=1}^{L_r}\sigma_{i}(r) \mbox{\;\;\;\;\; and
\;\;\;\;\;} 
\mathrm{VB}_\delta(r)=\frac{1}{L_r}\sum_{i=1}^{L_r}\epsilon_{i}(r),
\end{equation} 
where $L_r$ is the number of windows of length $r$. This is repeated over a
range of window sizes. Both quantities follow a power-law behaviour for
self-similar time series (Schmittbuhl et al. 1995) such that
$\mathrm{VB}_{w}(r)=r^H $ and $\mathrm{VB}_{\delta}(r)=r^H $. Thus, the
self-similarity exponent, $H$, is obtained from the slope of the corresponding
log-log plot.
\subsection{Wavelets}
A wavelet $\psi$ is a function with zero average and is normalised to one. A
family of wavelets is generated by scaling $\psi$ by a factor $s$ and
translating it by $u$ ($ \psi_{u,s}(t) =\frac{1}{\sqrt{s}} \psi \left( \frac
{t-u}{s} \right) $). The wavelet transform allows to construct a
time-frequency representation of a signal, the wavelet spectrum. One can then
infer the self-similarity parameter from the wavelet spectrum via ordinary
least squares at large wavelet scales (Abry and Veitch 1998, Stoev and Taqqu
2005).
\subsection{Rescaled Range}
The rescaled range $R/S$ (Hurst 1965) is a technique for estimating the LRD
parameter $d$ from a time series. The $R/S$ estimator is given by
\begin{equation}
R/S(\tau) = \frac {\max x(t,\tau) - \min x(t,\tau)} {\sqrt{ \frac{1}{\tau}
\sum_{t=1}^{\tau} ({\xi(t) - \langle \xi \rangle_\tau})^2 }},
\end{equation}
where $1 \leq t \leq \tau$, $ x (t,\tau) = \sum_{u=1}^{t} {x(u) - \langle x
\rangle_\tau} $, $ \langle x \rangle_\tau = \frac{1}{\tau} \sum_{t=1}^{\tau}
x(t) $. It scales like $\tau^J$ where the value of $J$ can be estimated from
the slope of $R/S$ from a log-log plot. 
\subsection{Detrended Fluctuation Analysis}
Detrended Fluctuation Analysis (DFA; Peng et al. 1994) also estimates the LRD
parameter $d$ from a time series. In DFA, a
profile is first computed by $ Y(i)=\sum_{t=1}^{N} x(t) $. The profile is cut
into $N_s$ non-overlapping segments of equal length$\ s$ and then the local
trend is subtracted for each segment $v$ by a polynomial least-squares fit of
the data. Linear (DFA1), quadratic (DFA2), cubic (DFA3) or higher order
polynomials can be used for detrending. In the $n^{th}$ order DFA, trends of
order $n$ in the profile, and of order $n-1$ in the original record, are
eliminated. Next the variance for each of the$\ N_s$ segments is
calculated by averaging over all data points$\ i$ in the$\ v^{th}$ segment:
\begin{equation}
F_s^2(v)=\langle Y_s^2(i) \rangle=\frac{1}{s}\sum_{i=1}^{s} Y_s^2 \left [
(v-1)s+i \right ]. 
\end{equation} 
Finally, the average over all segments is computed and the square root is
applied to obtain the following {\it fluctuation function}  
\begin{equation}
F(s)=\sqrt{\frac{1}{N_s}\sum_{v=1}^{N_s} F_s^2(v)}.
\end{equation}   
For different detrending orders, $n$, we obtain different fluctuation
functions $F(s)$, which are denoted by $F^{(n)}(s)$. The fluctuation function
scales according to $ F^{(n)}(s) \sim s^\zeta $, with $ d = \zeta - \frac {1}
{2}$. There are many variants of DFA, but use of standard DFA is
recommended by Basahn et al. (2008) if the functional form of a trend is not
a priori known. 
\subsection{Exact Whittle Estimator}
The Exact Whittle estimator is a semi-parametric estimator (Shimotsu and
Phillips 2005, 2006). This method assumes that the underlying model of LRD can
be represented by $ (1-B)^d X_t = \varepsilon_t $, where $ \varepsilon $ is
iid noise. See Shimotsu and Phillips (2005, 2006) for more details.
\subsection{Power Spectral Method}
As a semi-parametric estimator we use the power spectral method of Geweke
\& Porter-Hudak (1983) and Hurvich et al. (2001). Spectral methods find $d$ by
estimating the spectral slope. The periodogram is used, which is an
estimate of the spectral density of a finite-length time series and is given
by:
\begin{equation}
\hat{S}(\lambda_j)=\frac{1}{N}\left| \sum_{t=1}^{N} X(t)e^{-i2\pi t \lambda_j}
\right| ^2, \quad j=1,...,[N/2],
\end{equation} 
where$\ \lambda_j=j/N$ is the frequency and the square brackets denote
rounding towards zero. A series with LRD has a spectral density proportional
to $\left| \lambda \right|^{-2d} $ close to the origin. Since
$\hat{S}(\lambda) $ is an estimator of the spectral density, $d$ is estimated
by a regression of the logarithm of the periodogram versus the logarithm of
the frequency $\lambda $. Thus having calculated the spectral density estimate
$\hat{S}(\lambda) $, semi-parametric estimators fit a power law of the form $
f(\lambda,b,d)=b \left| \lambda \right|^{d} $, where $b$ is the scaling
factor.
\begin{figure}
\begin{center}
\begin{tabular}{ll}
\includegraphics[width=32pc]{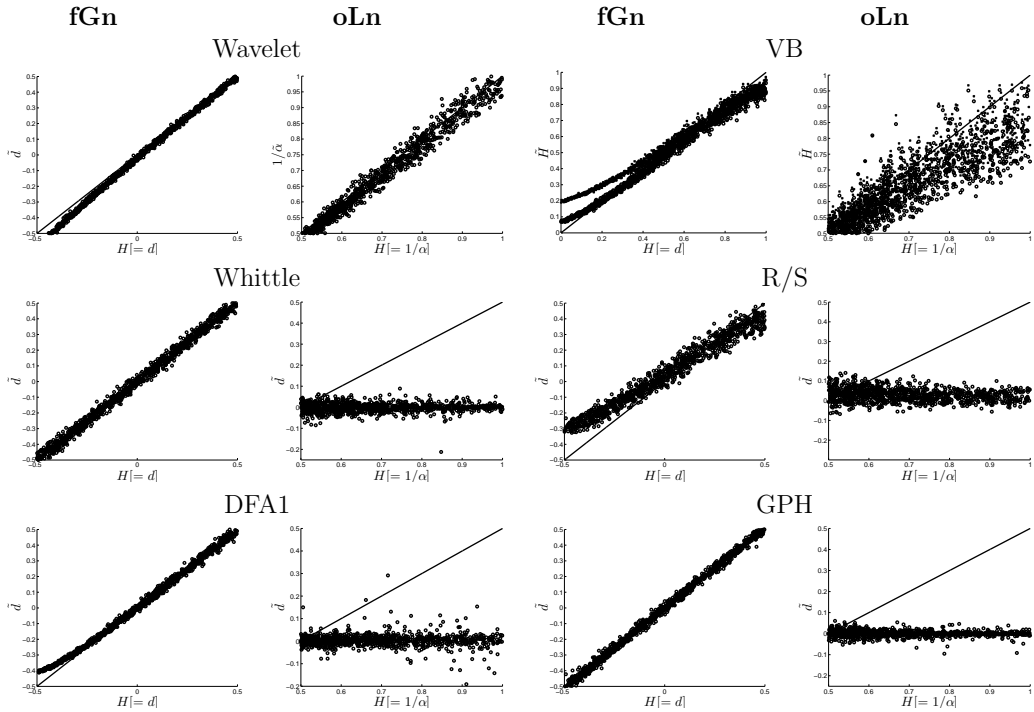} &
\end{tabular}
\caption{\label{fGnoLn}Estimates of $H$ for walks derived from fractional
  Gaussian noise (fGn) and ordinary Levy noise (oLn) for Wavelet method and
  the Variable Bandwidth estimator $\mathrm{VB}$ (crosses denote
  $\mathrm{VB}_\delta$ and circles $\mathrm{VB}_w$). Estimates of $d$ for fGn
  and oLn for Whittle estimator, $R/S$, DFA1 and GPH. The solid line indicates
  the true self-similarity parameter $H$ (d in the case of fGn or $1/\alpha$
  for oLn).}
\end{center}
\end{figure}

\section{Empirical Tests}  
In order to examine the statistical properties of the LRD estimators, like
bias, spread and outliers, we generate various test time series from the above
paradigmatic models. Specifically, we generate ensembles of 1000 members with
randomly selected parameters from uniform distributions: fGn with $ d \in
(-0.5,0.5) $, oLn with $ \alpha \in (1,2) $, ARFIMA with Gaussian increments
ARFIMA-G$(1,d,1)$ and with ordinary Levy increments ARFIMA-L$(1,d,1)$ with the
autoregressive coefficient $ a_1 \in (-\frac {1} {2}, \frac {1} {2}) $,
and the moving average coefficient $ b_1 \in (0,1) $, $ d\in (-\frac
{1} {2}, \frac {1} {2}) $, and $ \alpha \in (1,2) $. We also use various
first order autoregressive processes with parameters randomly chosen in
(-1,1). The time series length is always $2^{15} - M$ with $M=6000$ (see Stoev
\& Taqqu (2004) for an explanation of $M$) which is comparable to the length
of most observed climatic and other natural time series, thus long enough for
our purposes.
\subsection{Paradigmatic Time Series}
First we examine how well the SS and LRD estimators work for the models of
self-similarity. Most of the methods (e.g. DFA and $R/S$) require a regression
fit in order to estimate the  SS or LRD parameters. As shown by Chen et
al. (2002) non-stationarities and short-range dependencies can cause
crossovers in the fluctuation curves. Because of this we only regress on
the long-range part of the fluctuation curves. Our results are robust to the
particular choice of the cut off. 

\begin{figure}
\begin{center}
\begin{tabular}{ll}
\includegraphics[width=32pc]{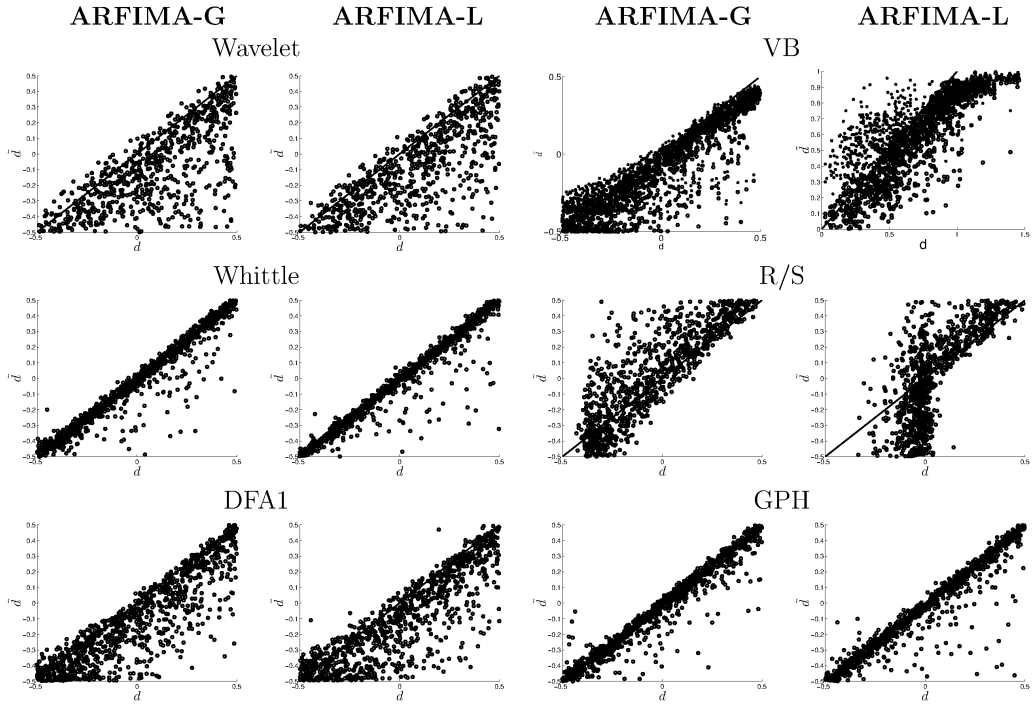} &
\end{tabular}
\caption{\label{ARFIMA2}Estimates of $H$ for walks derived from
  ARFIMA-G(1,d,1) and ARFIMA-L(1,d,1) with the Wavelet method and Variable
  Bandwidth estimator $\mathrm{VB}$. Estimates of $d$ for  ARFIMA-G(1,d,1) and
  ARFIMA-L(1,d,1) for the Whittle estimator, $R/S$, DFA1 and GPH. The solid
  line indicates the true self-similarity parameter $H$.}
\end{center}
\end{figure}

As can be seen in Fig. \ref{fGnoLn} the Wavelet and the $\mathrm{VB}$ methods
are the only methods which are able to infer the self-similarity of oLn. While
the Wavelet method has no large estimation spread, the $\mathrm{VB}$ method
exhibits with large errors. All other estimators estimate $d$ rather than $H$,
with $R/S$ producing the largest estimation variance, while Whittle, DFA1 and
GPH have considerably smaller estimation variances but with the odd outlier
(Fig. \ref{fGnoLn}). All four estimators do a reasonably good job of inferring
$H$ from fGn with $\mathrm{VB}_{w}(r)$ having the largest bias for $H$ close
to 0 and close to 1, and $R/S$ also having some bias at small $H$ with a
relatively large estimation variance. Wavelet, DFA1 and GPH produce very tight
estimates, with Wavelet and DFA1 only biased for small values, and Whittle and
GPH working well over the entire range. For these paradigmatic time series the
two semi-parametric estimators, Whittle and GPH, give the best estimates. This
picture changes once we allow for short-term dependence structures to
contaminate the pure self-similar character of fGn and oLn. As
Fig. \ref{ARFIMA2} shows all estimators work considerably worse for ARFIMA
surrogate data with large estimation spreads and many outliers. Again the
Whittle estimator and GPH perform better than the other estimators.

Before we go on to examine how well the estimators work with superimposed
trends, we test their accuracy on data generated from three basic short-range
dependence models: independent white noise, AR(1) and ARFIMA(1,0,1). As
Figs. \ref{PAR3}a, b and c show, Whittle estimator, GPH and DFA have the least
bias. But it also has to be noted that all estimators have many outliers,
suggesting that given an individual time series the estimates of $d$ or $H$
can be severely biased. The performance of higher order DFA is very similar to
DFA1 in the above test cases (not shown). These results are consistent with
previous empirical studies (e.g. Taqqu et al. 1995).
\subsection{Trends}
Another important test is to see how the presence of trends affects the
various estimators. Here we consider three cases: 1) a linear trend
superimposed by realisations of fGn, oLn, ARFIMA-G and ARFIMA-L; and 2) a
linear trend only in the second half of the time series superimposed by
realisations of fGn, oLn, ARFIMA-G and ARFIMA-L; 3) a linear trend in the
variance. Cases 2 and 3 are motivated by climate change where the time series
may (case 2) or may not (case 1) include the pre-industrial era, and where
climate change also influences the frequency and strength of fluctuations
(e.g. storms; case 3). Based on the evidence gathered so far, it is reasonable
to expect this can lead to bias any long-range dependence estimate.

\begin{figure} 
\begin{center} 
\begin{tabular}{ll}
\includegraphics[width=24pc]{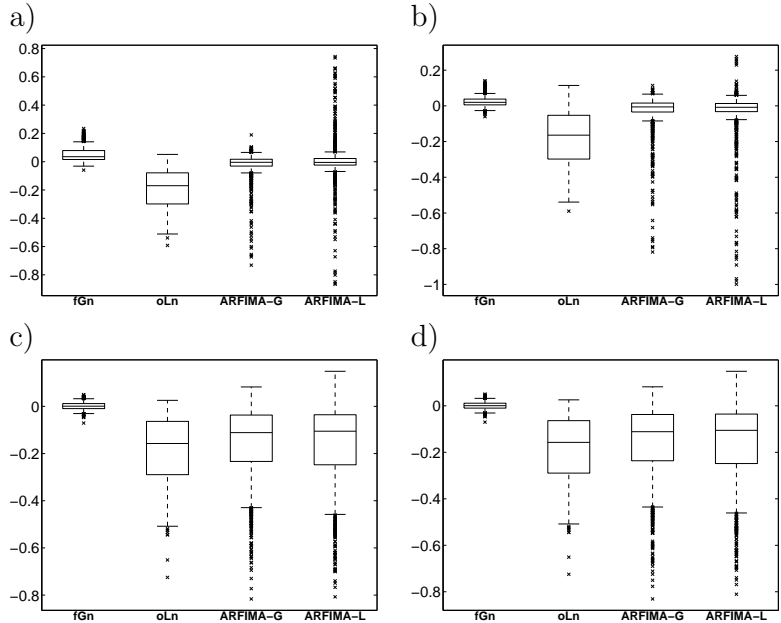} &
\end{tabular}
\caption{\label{TREND}Box plots of the difference between the nominal
  parameter value and the empirical parameter estimate a) GPH estimator for
  noise plus linear trend, b) GPH estimator for noise plus linear trend in
  second half of time series, c) DFA2 estimator for noise plus linear trend
  and d) DFA2 estimator for noise plus linear trend in second half of time
  series. On each box the central mark is the median, the edges of the boxes
  are the 25th and 75th percentiles, the whiskers extend to the most extreme
  data points not considered to be outliers and outliers are marked
  individually.}
\end{center}
\end{figure}

For cases 1 and 2, we assume the magnitude of the linear trend to be 1. The
empirical tests reveal that the GPH estimator is slightly biased for fGn,
ARFIMA-G and ARFIMA-L and has a relatively large negative bias for oLn
(Figs. \ref{TREND}). DFA is least biased for fGn data but has considerable
bias for oLn, ARFIMA-G and ARFIMA-L (Figs. \ref{TREND}) generated data. Both
ARFIMA-G and ARFIMA-L estimators have a large number of outliers. Both $R/S$
and $\mathrm{VB}$ show qualitatively similar behaviour compared to GPH (not
shown). Our results are qualitatively consistent with the study by Hu et
al. (2001).
\begin{figure}
\begin{center}
\begin{tabular}{ll}
\includegraphics[width=24pc]{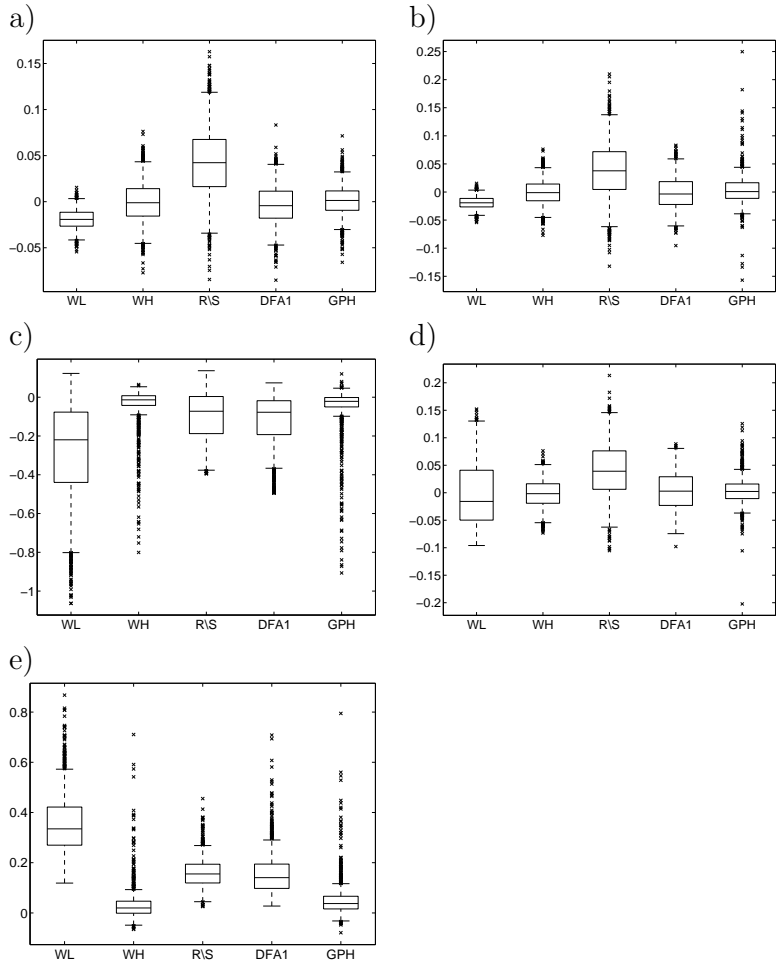} &
\end{tabular}
\caption{\label{PAR3}Box plots of the difference between the nominal parameter
  value and the empirical parameter estimate: a) uncorrelated Gaussian white
  noise, b) AR(1) and c) ARFIMA-G(1,0,1), d) AR(1) process with linear trend
  in variance and e) Markovian SDE with CAM noise. On each box the central
  mark is the median, the edges of the boxes are the 25th and 75th
  percentiles, the whiskers extend to the most extreme data points not
  considered to be outliers and outliers are marked individually.}
\end{center}
\end{figure}

Now we examine how the estimators handle a trend in the variance (case 3). For
this purpose we use an AR(1) model with increasing variance:
\begin{equation}
x_{t+1} = F + \alpha x_t + \sigma ( 1 + \frac {t} {10000} ) \zeta_t.
\end{equation}
Here, again, we generate 1000 realisations by sampling values for F from a
uniform distribution $ U(0,2) $, using $\alpha \in U(-0.5,0.5) $, and $ \sigma
\in U(0,1) $. For this case DFA is the least biased and GPH and $R/S$ show
considerable bias while the $\mathrm{VB}$ estimators have a huge bias,
although, their estimates are very narrow (Fig. \ref{PAR3}e).
\subsection{CAM noise}
Both LFSM and ARFIMA are additive noise models. In previous studies it has been
shown that multiplicative noise is important in natural systems (e.g. Majda et
al. 2009, Steinbrecher and Weyssow 2004). This raises the question of what the
effect of multiplicative noise would be on long-range dependence
estimators. To investigate, we use data generated from a process which has a
so called Correlated Additive and Multiplicative (CAM) noise term. An example
is the normal form for reduced climate models (Majda et al. 2009), which is
given by the following Stochastic Differential Equation (SDE):
\begin{equation} 
d x = (F + a x + b x^2 - c x^3)dt + \sigma ( L + I x ) dW. 
\label{CAM} 
\end{equation} 
As shown in Majda et al. (2009), the PDF of Eq. (\ref{CAM}) exhibits a
power-law decay over a particular range, although its tail ultimately decays
exponentially. We generate 1000 realisations by sampling random values for the
SDE parameters from an uniform distribution while complying with the parameter
relations as described in Majda et al. (2009).

The GPH estimator is slightly biased towards positive values while DFA and
$R/S$ have larger biases. Observe that the Wavelet estimator (as well as both
$\mathrm{VB}$ estimators (not shown)) estimate self-similar behaviour
(Fig. \ref{PAR3}f). While the PDF of Eq. (\ref{CAM}) decays in a power-law
like way over a given range it is not self-similar because its ultimate decay
is exponential (Majda et al. 2009). This makes the estimates obtained by the
wavelet and $\mathrm{VB}$ estimators questionable. Furthermore, $R/S$, DFA,
GPH and the Whittle estimator again show an uncomfortably large number of
outliers, again suggesting that the estimators may not be very reliable for
this case, or that the signal is not characterised simply by $H$ and $d$. We
note that Kantelhardt et al. (2002) have studied the performance of more
general multifractal DFA methods which were found to extract the full range of
scaling exponents in a particular multifractal test case.

\section{Natural Time Series Examples}
Now we return to the two natural time series from Section 2, and analyse their
self-similarity and LRD characteristics. We have shown above that the two time
series are non-Gaussian; are they also long-range dependent or self-similar? 

Applying the various estimation methods to the Faraday-Vernadsky temperature
time series gives evidence for long-range dependence but with a wide variety
of values: Whittle $d=0.24$, GPH $d=0.28$, DFA2 $d=0.43$, $R/S$ $d=0.33$,
Wavelet $H=0.53$, $\mathrm{VB}_w$ $H=0.92$ and $\mathrm{VB}_\delta$
$H=0.96$. While the three LRD estimators all provide evidence for long range
dependence, they provide a rather large range of values of the LRD parameter,
$d$. However, all of the estimates are between 0 and 0.5, suggesting a
long-range dependent but stationary process. The $H$ estimates are larger than
the $d$ estimates; this might suggest self-similarity with $ \alpha \approx
1.5 $.
There is also evidence for long-range dependence in the AE index, again with a
wide variety of estimates from the various estimators: Whittle $d=0.28$, GPH
$d=0.72$, DFA2 $d=0.66$, $R/S$ $d=0.4$, Wavelet $H=0.94$, $\mathrm{VB}_w$
$H=0.97$ and $\mathrm{VB}_\delta$ $H=0.99$. The estimates from GPH and DFA2
are larger than 0.5, suggesting a non-stationary time series, while the
Whittle and $R/S$ estimates suggests a stationary long-range dependent time
series. Again the $H$ estimates are larger than the $d$ estimates, providing
some evidence for self-similarity. Other evidence has been presented that AE
may not be a simple fractal (Consolini et al. 1996); recently work has focused
on high-frequency non-stationary and lower-frequency $1/f$ properties (Rypdal
and Rypdal 2010).

While all LRD estimators agree that there is evidence for LRD in these two
time series they provide a rather large range of possible values, illustrating
the problem of statistically robust LRD estimation in practice and the need
for further investigation of the performance of statistical indicators in the
presence of departures from fractality, including possible multifractality.

\section{Conclusions}
There are two contributions to self-similarity: (i) long-range dependence and
(ii) non-Gaussian jumps. This is not always appreciated in the various
communities with interests in detecting self-similarity and long-range
dependence. We have shown that empirical estimators of long-range dependence
are at best biased for fractional Gaussian noise, but at worst not robust for
processes which deviate from this idealised model. We have also shown that the
empirical estimators are not very robust in the presence of trends and
multiplicative noise.

Our results have several important implications for the modelling of natural
time series. In our view the ARFIMA model is a much better paradigmatic model
of natural time series than fGn since it allows one to explicitly model
short-range and long-range behaviour while also allowing for non-Gaussian
increments.

Finally, there is a need for estimation procedures which can deal with
multiplicative noise and trends in the variance. Such effects introduce
sizeable biases and estimation uncertainty. As such, all estimations of LRD
have to be taken with precaution. While it is true that all of the estimators
we tested perform reasonably well for fractional Gaussian noise, once a time
series is non-Gaussian or is non-stationary (in trend or volatility) the
estimators can be problematic. 

\begin{acknowledgements}
We thank M. Freeman and two anonymous reviewers for their comments on an
earlier version of this manuscript. In particular we are grateful to one of
the reviewers for their emphasis on the Whittle and wavelet estimators. This
study is part of the British Antarctic Survey Polar Science for Planet  Earth
Programme. It was funded by The Natural Environment Research Council. TG
acknowledges generous support through a EPSRC studentship. RBG was funded
under EPSRC research grant EP/D065704/1.
\end{acknowledgements}

\end{document}